\documentclass[aoas]{imsart}
\RequirePackage{amsthm,amsmath,amsfonts,amssymb}
\RequirePackage[authoryear]{natbib}
\usepackage[utf8]{inputenc}
\usepackage[skip=10pt]{caption}
\usepackage{bm}
\usepackage{graphicx}
\usepackage[pdfencoding=auto]{hyperref}
\usepackage{float}
\usepackage{color}
\usepackage{xcolor}
\usepackage{soul}
\usepackage{ulem}
\usepackage{array}
\usepackage[bottom,hang,flushmargin]{footmisc}

\begin{document}
\begin{frontmatter}

\title{Comparison of Combination Methods to Create Calibrated Ensemble Forecasts for Seasonal Influenza in the U.S.}

\begin{aug}

\author[A]{\fnms{Nutcha} \snm{Wattanachit}\ead[label=e1,mark]{nwattanachit@umass.edu}},
\author[A]{\fnms{Evan L.} \snm{Ray}},\\
\author[B]{\fnms{Thomas C.} \snm{McAndrew}}
\and
\author[A]{\fnms{Nicholas G.} \snm{Reich}},
%%%%%%%%%%%%%%%%%%%%%%%%%%%%%%%%%%%%%%%%%%%%%%
%% Addresses                                %%
%%%%%%%%%%%%%%%%%%%%%%%%%%%%%%%%%%%%%%%%%%%%%%
\address[A]{School of Public Health and Health Sciences, University of Massachusetts Amherst, \printead{e1}}
\address[B]{College of Health, Lehigh University}
\end{aug}
\begin{abstract}
The characteristics of influenza seasons varies substantially from year to year, posing challenges for public health preparation and response. Influenza forecasting is used to inform seasonal outbreak response, which can in turn potentially reduce the societal impact of an epidemic. The United States Centers for Disease Control and Prevention, in collaboration with external researchers, has run an annual prospective influenza forecasting exercise, known as the FluSight challenge. A subset of participating teams has worked together to produce a collaborative multi-model ensemble, the FluSight Network ensemble. Uniting theoretical results from the forecasting literature with domain-specific forecasts from influenza outbreaks, we applied parametric forecast combination methods that simultaneously optimize individual model weights and calibrate the ensemble via a beta transformation. We used the beta-transformed linear pool and the finite beta mixture model to produce ensemble forecasts retrospectively for the 2016/2017 to 2018/2019 influenza seasons in the U.S. We compared their performance to methods currently used in the FluSight challenge, namely the equally weighted linear pool and the linear pool. Ensemble forecasts produced from methods with a beta transformation were shown to outperform those from the equally weighted linear pool and the linear pool for all week-ahead targets across in the test seasons based on average log scores. We observed improvements in overall accuracy despite the beta-transformed linear pool or beta mixture methods' modest under-prediction across all targets and seasons. Combination techniques that explicitly adjust for known calibration issues in linear pooling should be considered to improve ensemble probabilistic scores in outbreak settings. 
\end{abstract}


\end{frontmatter}

\section{Introduction}

Seasonal influenza outbreaks pose public health challenges and cause a large morbidity and mortality burden worldwide. The United States Centers for Disease Control and Prevention (CDC) estimates there were 35.5 million cases of influenza, 490,600 influenza-related hospitalizations, and 34,200 deaths from influenza during the 2018–2019 influenza season in the US \citep{cdc_burden_19}. Influenza forecasting has become integral to public health decision making \citep{lutz_2019}. A forecasting model uses data to make projections of the future trajectory of an infectious disease target, such as cases, hospitalizations and deaths, and can provide uncertainty measures of its predictions. Thus, forecasting models are a powerful tool for public health officials to improve seasonal outbreak preparedness and response, which can in turn potentially reduce the burden of seasonal influenza. The CDC's establishment of the Center for Forecasting and Outbreak Analytics in August of 2021 \citep{cdc_2021_center} highlights a critical need to advance the use of infectious disease forecasting and modeling.

To provide public health officials real-time, prospective information about the future trajectory of seasonal influenza, the CDC, in collaboration with external researchers, started an annual prospective influenza forecasting exercise in the U.S., known as the FluSight challenge, in 2013. This exercise has been conducted with the goal of improving forecast accuracy and the integration of forecasts with real-time public health decision making. Multiple academic and non-academic groups submit weekly forecasts to the FluSight challenge. A submission typically contains probabilistic and point forecasts for seven targets in each of the 10 Health and Human Services (HHS) regions in the U.S. as well as at the national level. In this manuscript, we focus on the probabilistic forecasts, in which a predictive distribution is specified for the outcome of interest. All forecast targets are based on the weighted percentage of outpatient visits for influenza-like illness (wILI) collected through the U.S. Outpatient Influenza-like Illness Surveillance Network (ILINet), weighted by state populations. 

Constructing a single ensemble forecast that combines the forecasts from multiple individual models has advantages. A ensemble forecast unifies signals from many models into a single forecast, making it easier for stakeholders to understand. In addition, ensemble forecasts have been shown to consistently achieve a high degree of accuracy and often outperform individual forecasts of infectious disease targets \citep{yamana_2016,yamana2017,defelice2017,viboud_2018,ray2020,chowell2020,cramer_2021,oidtman_2021,ray_2022}. A subset of teams participating in the FluSight challenge has produced a collaborative multi-model ensemble, the FluSight Network ensemble, using stacked generalization \textemdash in particular, the FluSight Network ensemble is calculated as a linear combination of the individual forecasts. 

Despite the success of linear combination methods such as the one used to produce the FluSight Network ensemble, their forecasts lack calibration \citep{hora2004,gneiting2013}. \cite{gneiting2013} proved that the linear aggregation increases the dispersion of the combined predictive distribution and therefore may result in overdispersed ensemble forecasts even when the individual forecasts are well-calibrated. More generally, a simple linear combination of individual forecasts may produce miscalibrated ensemble forecasts unless their calibration is adjusted for. 

Previous work has presented parametric and nonparametric approaches to combining and calibrating ensemble forecasts. The beta-transformed linear pool is a combination formula that calibrates the combined predictive distribution by overlaying the linear pool with a beta transformation \citep{ranjan2010,gneiting2013}. \cite{bassetti2018} proposed an extension to the beta-transformed linear pool, using a Bayesian nonparametric approach to estimate infinite beta mixture models. This method achieves a theoretically stronger result of probabilistic calibration compared to the beta-transformed linear pool by extending the flexibility of the combination function. \cite{kuleshov_21} introduced calibrated risk minimization as a principle that maximizes sharpness subject to calibration by adding calibration loss as a constraint in the loss function. \cite{rumack_21} presented a post-processing method called the recalibration ensemble that combines and calibrates forecasts in separate steps and applied this method to recalibrating epidemic forecasts.

In practice, there is merit in selecting parsimonious models and combination methods with computationally efficient estimation \citep{claeskens2016,baran2018,stanescu2016}. The optimal degree of flexibility and computational complexity of combination methods often vary for different applications. \cite{baran2018} compare the performance of multiple forecast combination methods and assesses the degree of flexibility combination methods needed to yield the best practical results for post-processing applications in forecasting wind speed and precipitation. In influenza probabilistic forecasting, \cite{ray2018_ens} study a range of weighing schemes with different levels of complexity in generating ensemble forecasts via the feature-weighted ensemble approach that combines aspects of linear pooling or stacking and gradient boosting. In both of these studies, the methods with an intermediate level of flexibility yielded better predictive performances in their respective applications.  

This work aims to add to the growing field of infectious disease probabilistic forecasting by investigating the accuracy and probabilistic calibration of ensemble forecasts produced from combination methods that combine and calibrate simultaneously while not having any knowledge of the underlying model structure of the individual models or the ability to reproduce their forecasts in the U.S. seasonal influenza setting. Using 27 individual models from the FluSight network, we apply the linear pool, beta-transformed linear pool, and the finite beta mixture approach to combine predictive distributions. We also examine whether more parsimonious approaches to the aforementioned methods with fixed, equal individual model weights are sufficient to produce accurate and well-calibrated ensemble forecasts. We modify estimation approaches of the methods with beta transformation to accommodate the binned probability distribution representation used in the FluSight challenge \citep{mcgowan2019}. 

Section \ref{sec2} reviews the CDC influenza data, forecast targets, and forecast combination methods. Section \ref{3} describes the application of the combination methods in seasonal influenza forecasting and presents results. Section \ref{4} contains discussions of results in the context of related work, real-time forecasting operations, and data-driven public health decision making.

\section{Methods}\label{sec2}

\subsection{Influenza Data}

\begin{figure}[ht]
\includegraphics[scale=0.62]{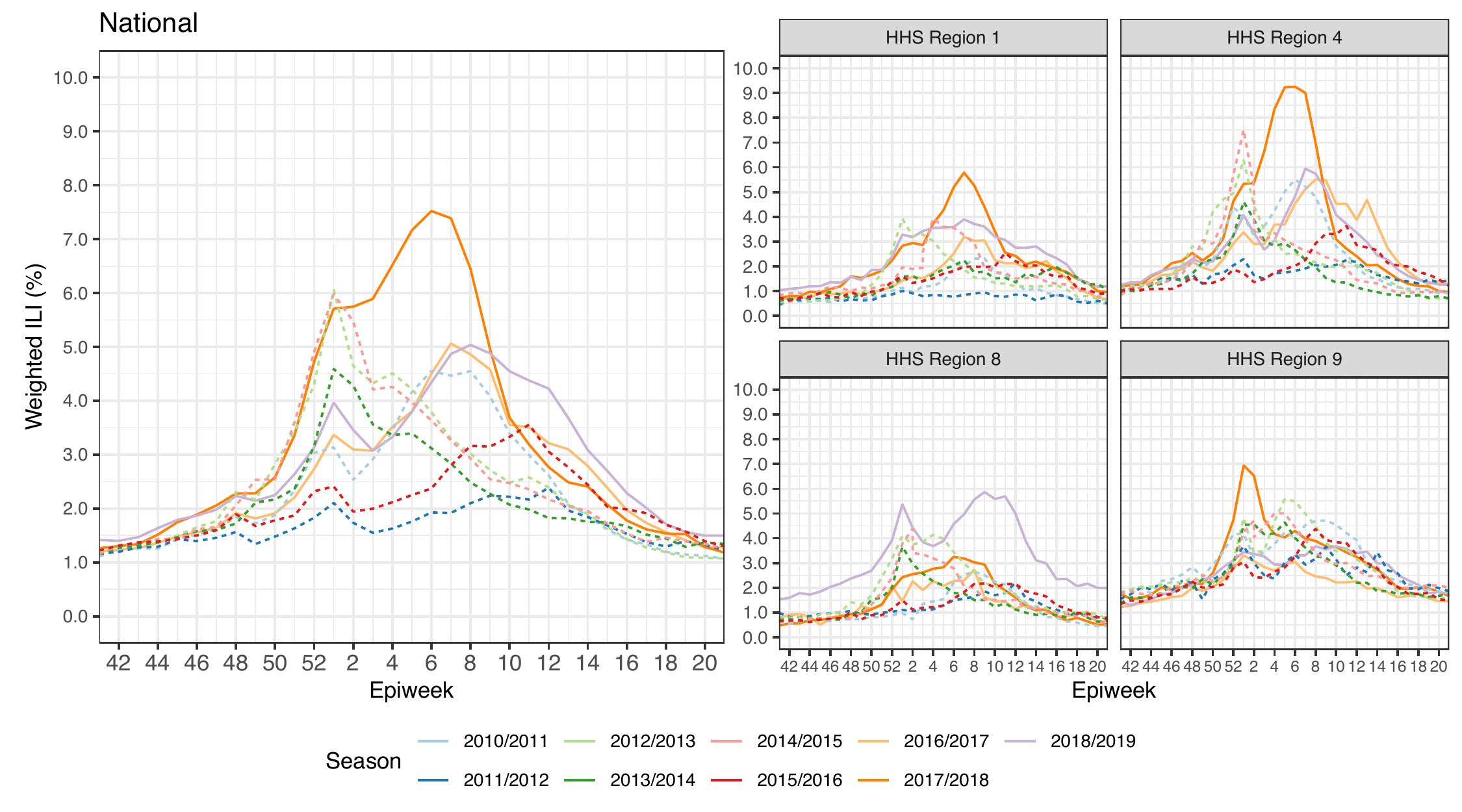}
\caption{Influenza-like illness weighted by state populations (wILI) data at a national level (left panel) and four HHS regions (right panel) from the 2010/2011 to 2018/2019 influenza season published by the U.S. Outpatient Influenza-like Illness Surveillance Network (ILINet). The 2016/2017 to 2018/2019 seasons, which are the test seasons, are represented in solid lines.}
\label{fig:wilidat}
\end{figure}

The U.S. Outpatient Influenza-like Illness Surveillance Network (ILINet) publishes the weekly percentage of outpatient doctor's office visits due to influenza-like illness weighted by state populations (wILI). ILINet is a syndromic surveillance system that includes more than 3,000 providers \citep{cdc_2020}. The CDC Influenza Division reports weekly estimates of wILI for the United States and for the 10 Health and Human Services (HHS) regions (\autoref{fig:wilidat}). 

\subsection{Forecast targets}

Forecasts submitted to the CDC FluSight challenge typically consists of three seasonal targets and four short-term targets. We produce ensemble forecasts of short-term 1-4 week ahead wILI for all locations from the 2016/2017 to 2018/2019 influenza season in this study. We do not include forecasts of seasonal targets, such as the peak week, peak incidence, and seasonal onset, due to the lack of importance of probabilistic forecasts after those events have been observed in a particular season.

\subsection{Forecast combination methods}

Let $f_1,...,f_M$ and $F_1,...,F_M$  be predictive probability density functions (PDFs) and cumulative distribution functions (CDFs), respectively, for a real-valued forecast target, $y$, from $M$ individual models. The combination methods described in this section include the linear pool as a baseline method and the beta-transformed linear pool and finite beta mixture combination as the methods that combine and calibrate forecasts. 

\subsubsection{Linear pool (LP and EW-LP)}
The linear pool is a mixture model with a predictive density 
\begin{equation}
f_{\text{LP}}(y)=\sum_{m=1}^M \omega_mf_m(y),
\end{equation}
where $\omega_m$ is a nonnegative weight for the $m^{\text{th}}$ individual model and $\sum_{m=1}^M \omega_m=1$. The equally weighted linear pool (EW-LP) is a special case of the LP with the weights fixed to $\omega_m=\frac{1}{M}$. 

\subsubsection{Beta-transformed linear pool (BLP and EW-BLP)}
\begin{figure}[ht]
\includegraphics[scale=0.22]{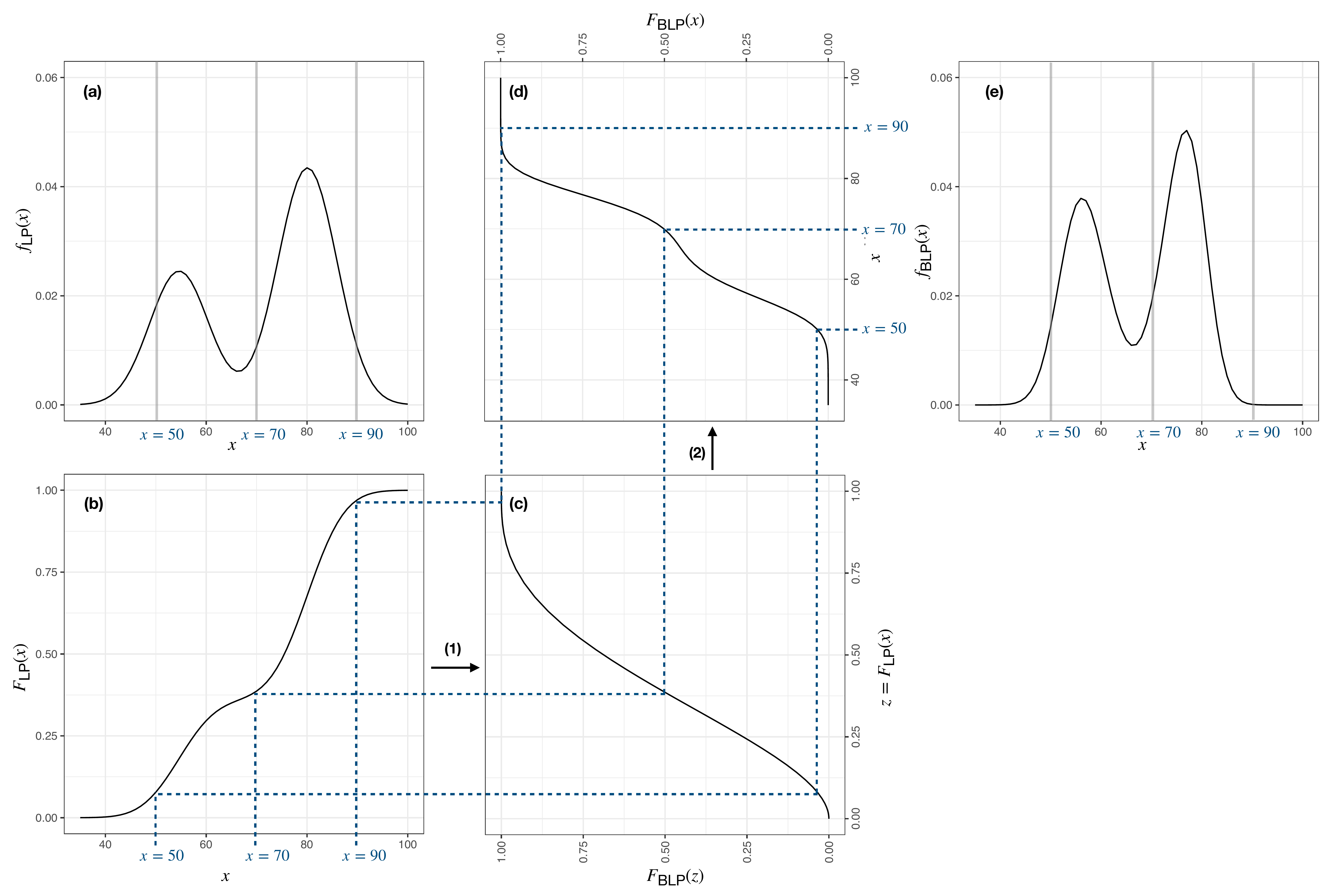}
\caption{Illustrative example of the BLP's beta transformation of $F_\text{LP}(x)$. For demonstration purposes, the individual model weights of the LP and BLP are fixed to be the same. We start with a predictive density from a linear pool, $f_\text{LP}(x)$ in panel (a), and the corresponding CDF $F_\text{BLP}(x)$ in panel (b). Step (1) shows the beta transformation of $z=F_\text{LP}(x)$ in panel (b) through $F_\text{BLP}(z)$ = $\text{B}_{\alpha,\beta}(z)$ with $\alpha=2$ and $\beta=3$ in panel (c). Step (2) shows the BLP's predictive CDF $F_\text{BLP}(x)$, panel (d), as a result of the beta transformation in the first step. Panel (d) shows that the beta-transformed CDF concentrates probability closer to the median and its predictive density, panel (e), is narrower compared to that of the LP in panel (a). Other choices of the parameters for the beta transform could lead to a still narrower distribution after the transformation, a wider distribution, or an asymmetric adjustment that acts differently in the left and right tails.}
\label{fig:transform}
\end{figure}

\cite{gneiting2013} demonstrate that the LP produces forecasts that lack calibration when the component forecasts are well-calibrated and propose a flexible alternative approach, the beta-transformed linear pool (BLP), which has a predictive CDF defined by
\begin{align}
F_{\text{BLP}}(y)=B_{\alpha,\beta}\Big(\sum_{m=1}^M \omega_m F_m(y)\Big),
\end{align}
where $B_{\alpha,\beta}$ denotes the CDF of the beta distribution with the parameters $\alpha,\beta > 0$, $\omega_m$ is a nonnegative weight for the $m^{\text{th}}$ individual model weight and $\sum_{m=1}^M \omega_m=1$. To find the predictive PDF of the BLP we can differntiate the above CDF, finding
\begin{equation}
f_{\text{BLP}}(y)=\Big(\sum_{m=1}^M \omega_mf_m(y)\Big)b_{\alpha,\beta}\Big(\sum_{m=1}^M \omega_m F_m(y)\Big)
\label{eq:blp-pdf}
\end{equation}
where $b_{\alpha,\beta}$ is the PDF of the beta distribution. The LP is a special case of the BLP when $\alpha=\beta=1$. The equally weighted component variation of this method is the equally weighted beta-transformed linear pool (EW-BLP), which is a special case of the BLP with fixed weights $\omega_m=\frac{1}{M}$. \autoref{fig:transform} demonstrates how the BLP's beta transformation operates on the LP's predictive CDF.

\subsubsection{Finite beta mixture combination (\texorpdfstring{$\text{BMC}_K$}{BMCK} and \texorpdfstring{$\text{EW-BMC}_K$)}{EW-BMCK}}

\cite{bassetti2018} use a Bayesian approach to extend the BLP to finite and infinite beta mixtures for combining and calibrating predictive distributions. \cite{baran2018} note the high computational costs of the estimating this approach. Due to the computational burden of this Bayesian approach, we choose to employ a frequentist approach to estimate a finite beta mixture model 
\begin{align}
F_{\text{BMC}_K}(y)=\sum_{k=1}^K \theta_kB_{\alpha_k,\beta_k}\Big(\sum_{m=1}^M \omega_{km} F_m(y)\Big),
\end{align} 
where $K$ is the number of beta components, $\theta_k$ is a beta mixture weight for the $k^{\text{th}}$ beta component, $B_{\alpha_k,\beta_k}$ denotes the CDF of the beta distribution with the parameters $\alpha_k,\beta_k >0$, and $\boldsymbol{\omega}_k=(\omega_{k1},..., \omega_{kM})$ comprises the individual model weights specific to each beta component. Differentiating the CDF, the predictive density of the $\text{BMC}_K$ is
\begin{equation}\label{eq:bmc-pdf}
f_{\text{BMC}_K}(y)=\sum_{k=1}^K \theta_k\Big(\sum_{m=1}^M \omega_{km} f_m(y)\Big)b_{\alpha,\beta}\Big(\sum_{m=1}^M \omega_{km} F_m(y)\Big).
\end{equation} 
The equally weighted variation of the finite beta mixture combination approach ($\text{EW-BMC}_K$) is a special case of the $\text{BMC}_K$ with $\boldsymbol{\omega}_k=(\frac{1}{M},...,\frac{1}{M})$ 
With $K=1$, the $\text{BMC}_K$ and the $\text{EW-BMC}_K$ become the BLP and the EW-BLP, respectively.

\subsection{Modification of the BLP and \texorpdfstring{$\text{BMC}_K$}{BMCK} methods for combining discrete distributions}
The predictive density functions of the BLP and the $\text{BMC}_K$ are given by the equation \eqref{eq:blp-pdf} and \eqref{eq:bmc-pdf}, respectively. However, the forecasts of 1-4 week ahead wILI, which is a  continuous measure of disease incidence, are represented using a binned probability format in submissions to the FluSight challenge. Here we describe a modification to BLP and $\text{BMC}_K$ models to handle this discretized representation of the target variable.\\ 
Let $F$ denote a predictive CDF of a forecasting model, $Y$ be the outcome variable, and   $\{(\ell_i, u_i]: i = 1, \ldots, I\}$ be a collection of disjoint bins covering the set of possible outcomes for $Y$, with $u_i = \ell_{i + 1}$ for $i < I$\footnote{The FluSight Challenge uses a slightly different binned probability format where the $i^{\text{th}}$ bin is defined as $[\ell_i,u_i)$ \citep{mcgowan2019}; this detail does not have a practical impact on the set up because the influenza-like-illness measure is continuous.}. An individual forecast of $Y$ consists of an assignment of probabilities to each of the $I$ bins:

\begin{align}
P_i & =Pr(\ell_i < Y \le u_i)  \\
    & =F(u_i)-F(\ell_i).
\end{align}

In order to estimate the parameters of the $\text{BMC}_K$, we modify the log-likelihood function, which is the log of equation \eqref{eq:bmc-pdf}, for a single observation $y$ that falls in bin $j$ to be 

{\small
\begin{align*}
\log\left[f_{\text{BMC}_K}(y)\right]
&=\log\left[P_{\text{BMC}_K,j}\right]\\
&=\log[F_{\text{BMC}_K}(u_j)-F_{\text{BMC}_K}(\ell_j)]\\
&= \log\left[\sum_{k=1}^K \theta_k B_{\alpha_k, \beta_k}\Big(\sum_{m=1}^M \omega_{km} F_{m}(u_j) \Big)-\sum_{k=1}^K \theta_k B_{\alpha_k, \beta_k}\Big(\sum_{m=1}^M \omega_{km} F_{m}(\ell_j)\Big)\right] \\
&= \log\left[\sum_{k=1}^K \theta_k B_{\alpha_k, \beta_k}\Big(\sum_{m=1}^M \omega_{km} \sum_{i \leq j}P_{m,i} \Big)-\sum_{k=1}^K \theta_k B_{\alpha_k, \beta_k}\Big(\sum_{m=1}^M \omega_{km} \sum_{i < j}P_{m,i}\Big)\right] 
\end{align*}
}%

where $P_{\text{BMC}_K,j}$ is the probability assigned to bin $j$ by the $\text{BMC}_K$'s discretized predictive distribution, $F_{\text{BMC}_K}(y)$ is the continuous predictive CDF of the $\text{BMC}_K$, $F_{m}(u_i)$ and $F_{m}(\ell_i)$ are the predictive CDFs of a individual model $m$, and $P_{m,i}$ is the probability assigned to bin $i$ by individual model $m$'s discretized predictive distribution.

Since the BLP is a special case of the $\text{BMC}_K$ where $K=1$, the modified log-likelihood function of the BLP is the same as above with a single beta component term in the outer summation. 

\subsection{Data and code accessibility}
All individual forecasts, data, and code used for conducting analyses presented in this manuscript are publicly available \citep{flusight,code_git}.

\section{Application in seasonal influenza forecasting in the U.S.}\label{3}

We apply the combination methods introduced in Section \ref{sec2} to prospective forecasts from 27 individual forecasting models (Table S1, Supplementary Material \citep{Wattana22}) available in the FluSight Network repository \citep{flusight} to generate weekly ensemble forecasts of 1-4 week ahead wILI for the United States and the 10 Health and Human Services (HHS) regions from the 2016/2017 to 2018/2019 influenza seasons. 

\subsection{Forecast evaluation}

We follow the FluSight Challenge guidelines \citep{cdc_1920} by using the logarithmic score or log score which is defined as the logarithm of the predictive density or mass function evaluated at the observed data point. The log score is a proper scoring rule that assesses the sharpness and calibration of probabilistic forecasts simultaneously \citep{gneiting2007}.
In the FluSight challenge where forecasts are represented in a binned probability format, the log score is defined as
\begin{align*}
\text{LogS}(f,y^*)&=\log\int^{u_i}_{\ell_i} f(y)dy \\
&=\log P_i,
\end{align*}
where $y^*$ is the observed value of the forecast target $y$ and $\ell_i$ and $u_i$ are the pre-specified lower and upper bounds of bin $i$ such that $y^*\in[\ell_i,u_i)$. 

We generate forecasts from each combination method for all combinations of week, region, target, and season, and calculate their log scores. Following the CDC scoring convention \citep{mcgowan2019}, we truncate log scores to be no lower than $-10$. The benefit of this approach is that it enables us to average log scores for a method even when that method receives a log score of $-\infty$ (assigning zero probability to an observed value) for any forecasts. However, this modified log score is formally no longer a proper score. Log scores are averaged across all forecast regions and weeks for each target and test season to get summary measures of accuracy for each method to compare their performance. 

The calibration of a probabilistic forecast addresses the statistical consistency between the predictive distributions of forecasts and the observations. The concept of calibration allows us to assess whether a model produces reliable forecasts, i.e. whether an event that the model assigns a particular predicted probability really occurs that at that frequency in the long run. The forecast $f$ is probabilistically calibrated if its probability integral transform (PIT) values are uniformly distributed on the unit interval \citep{gneiting_calib_07}. The probability integral transform, $z_i=P_i(y^*)$, is the probability obtained from evaluating the predictive CDF of a model at the observed value ($y^*$)\footnote{Since the target variables are discretized in this application, we adapt Definition 2.5 in \cite{gneiting2007} by sampling a uniformly distributed PIT value between $F(l_i)$ and $F(u_i)$ where $[l_i, u_i)$ is the bin containing the observed value.}.

To assess the probabilistic calibration of the ensemble forecasts (i.e., the uniformity of the PIT values), we use the graphical tool called the probability plot, which plots the empirical CDF of the PIT values. Specifically, we compute the PIT values of all observations in the test seasons and plot their the empirical CDFs by target and season. The empirical CDF curve should follow a 45-degree line bisecting the plot if the forecasts are probabilistically calibrated. In the case where deviations from uniformity are observed, the shape of empirical CDF curve of the PIT values suggests the causes behind the lack of probabilistic calibration \citep{laio2007}. For example, PIT values concentrating near 0 and 1 indicates that the observed values fall on the tails of the predictive distribution of the forecasts more frequently than expected, i.e., the probability plot shows the slope steeper than 1 near the PIT values of 0 and 1, so that the predictive distributions were too narrow. To quantitatively measure the deviation of a PIT CDF curve from a standard uniform CDF, we compute the Cramer distance \citep{gneiting2007,rizzo_16}, $\int^\infty_{-\infty}(F(x)-G(x))^2dx$, where $F(x)$ is a PIT CDF curve and $G(x)$ is a standard uniform CDF. The Cramer distance can be viewed as a summary measure of calibration, however, it lacks the diagnostic property of the probability plot.

\subsection{Parameter estimation}

In our application, the training data set consists of individual forecasts and influenza data from the 2010/2011 to 2018/2019 season. The test data set includes the 2016/2017 to 2018/2019 influenza seasons. When generating forecasts for a test set season, the training data set consists of all the influenza seasons preceding that season.

The parameters of each combination method are estimated simultaneously by maximizing the average log score, which is positively oriented (i.e., higher scores are better), via maximum likelihood estimation over a training data set. The parameters are chosen to be target-specific to allow for variations among targets. We update the parameter estimates for each test season, so there are 12 sets of parameters to be estimated for four targets and three test seasons. We modify the estimation approach for the BLP, EW-BLP, $\text{BMC}_K$, and $\text{EW-BMC}_K$ as outlined in the Methods section in order to apply the combination methods to individual forecasts in a binned probability representation. 

\subsubsection{Choice of \texorpdfstring{$K$}{K} for finite beta mixture combination approaches}

We use a leave-one-season-out cross validation approach to select the number of beta components, $K$, in the $\text{BMC}_K$ and $\text{EW-BMC}_K$ for each target-test season pair. Specifically, we train the $\text{BMC}_K$ and $\text{EW-BMC}_K$ using $K=2$ through 5 on each subset of data in the training data with one season left out and use those ensemble fits to generate forecasts for the left out influenza season. Log scores for all combination methods are calculated for all unique forecasts, then averaged across all weeks, regions, and validation seasons to obtain a single mean validation log score for each target and method. In order to take model complexity into account, we calculate mean validation log scores across all locations for each validation season in training seasons, compute a standard error for each target-test season pair, and select the smallest $K$ for $\text{BMC}_K$ and $\text{EW-BMC}_K$ with mean validation log scores within 1 standard error of the best log score in a particular target-test season pair. 

Based on the mean validation log scores in \autoref{tab:cvls}, $K=2$ is selected for the $\text{BMC}_K$ and $\text{EW-BMC}_K$ for all four targets and three seasons. The variation with $K = 2$ has the best mean validation log scores in every instance other than the 2-week ahead target in the 2018/2019 season. Overall, using a higher number of beta components in the finite beta mixture approaches does not substantially improve mean out-of-sample log scores in our application. Thus, the finite beta mixture methods with the most parsimonious number of parameters are selected.

\begin{table}[ht]
\footnotesize
\centering
\caption{Mean validation log scores of $\text{BMC}_K$ and $\text{EW-BMC}_K$ for all target-season pairs. Scores shown in bold are the selected methods.}
\label{tab:cvls}
\begin{tabular}{@{}ccccccccc@{}}
\multicolumn{9}{ l }{2016/2017}\\
\hline
Week\\
ahead& \multicolumn{1}{c}{$\text{BMC}_2$}& \multicolumn{1}{c}{$\text{BMC}_3$}& \multicolumn{1}{c}{$\text{BMC}_4$}& \multicolumn{1}{c}{$\text{BMC}_5$}& \multicolumn{1}{c}{$\text{EW-BMC}_2$}& \multicolumn{1}{c}{$\text{EW-BMC}_3$}& \multicolumn{1}{c}{$\text{EW-BMC}_4$}& \multicolumn{1}{c}{$\text{EW-BMC}_5$}\\
\hline
1  & \bf{-2.49} & -2.50 & -2.50 & -2.50 & \bf{-2.50} & -2.50 & -2.50 & -2.50\\
2  & \bf{-2.74} & -2.75 & -2.75 & -2.76 & \bf{-2.76} & -2.76 & -2.76 & -2.76\\
3  & \bf{-2.95} & -2.95 & -2.95 & -2.97 & \bf{-2.92} & -2.92 & -2.92 & -2.92\\
4  & \bf{-3.08} & -3.09 & -3.10 & -3.11 & \bf{-3.03} & -3.03 & -3.04 & -3.04\\
\hline
\multicolumn{9}{ l }{2017/2018}\\
\hline
1  & \bf{-2.49} & -2.50 & -2.50 & -2.51 & \bf{-2.51} & -2.51 & -2.51 & -2.52\\
2  & \bf{-2.75} & -2.75 & -2.76 & -2.77 & \bf{-2.78} & -2.78 & -2.78 & -2.78\\
3  & \bf{-2.96} & -2.96 & -2.97 & -2.98 & \bf{-2.94} & -2.95 & -2.95 & -2.95\\
4 & \bf{-3.09} & -3.09 & -3.10 & -3.10 & \bf{-3.06} & -3.06 & -3.06 & -3.06\\
\hline
\multicolumn{9}{ l }{2018/2019}\\
\hline
1  & \bf{-2.51} & -2.52 & -2.52 & -2.53 & \bf{-2.55} & -2.55 & -2.55 & -2.55\\
2  & \bf{-2.80} & -2.79 & -2.81 & -2.80 & \bf{-2.85} & -2.84 & -2.85 & -2.85\\
3  & \bf{-2.99} & -3.00 & -3.00 & -3.01 & \bf{-3.03} & -3.03 & -3.03 & -3.03\\
4  & \bf{-3.13} & -3.14 & -3.13 & -3.14 & \bf{-3.15} & -3.15 & -3.15 & -3.15\\
\hline
\end{tabular}
\end{table}

\subsection{Results}

\subsubsection{Overall Summary}

Based on mean out-of-sample log scores across all targets and seasons (\autoref{fig:ls_comb} panel (b)), the $\text{BMC}_2$ is the most accurate method, followed by the BLP and LP. Across all three test seasons, the $\text{BMC}_2$ outperformed the other five methods for 3 and 4 week ahead horizons, and performed as equally well as the BLP for the 2 week ahead horizon (\autoref{fig:ls_comb} panel (a)). The BLP is the best performing method for the 1 week ahead horizon. The $\text{BMC}_2$ is also the best performing method for the 2017/2018 and 2018/2019 season based on mean out-of-sample log scores across all four horizons, while the BLP is the best performing method for the 2016/2017 season. These results indicate that the BLP and $\text{BMC}_2$ can consistently improve the accuracy of ensemble forecasts compared to the other commonly used methods included in this study despite season-to-season and target variations.

The 1 week ahead forecasts from the BLP and $\text{BMC}_2$ methods are more probabilistically calibrated than other methods based on the probability plots and their Cramer distances from the standard uniform CDF. However, the forecasts produced from all beta-transformed methods became less calibrated as forecast horizons increased. 

\subsubsection{Comparison of combination methods' accuracy}

\begin{figure}[htp]
\includegraphics[scale=0.9]{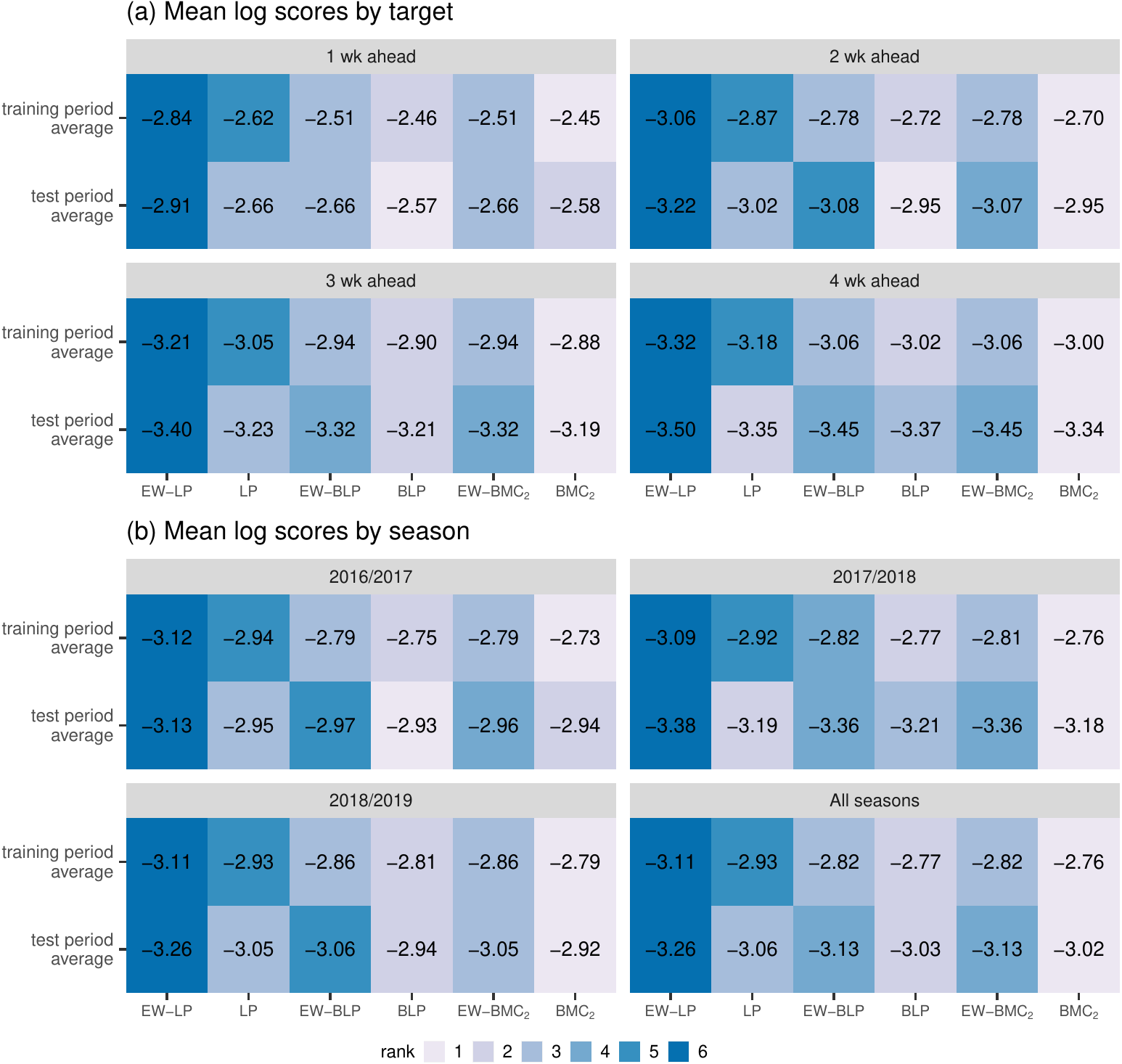}
\caption{Mean training and out-of-sample log scores of ensemble forecasts of wLIL in the U.S. Panel (a) shows mean training and out-of-sample log scores by target. Panel (b) shows mean training and out-of-sample log scores by season and across all seasons. Higher log scores (lower ranks) indicate better accuracy.}
\label{fig:ls_comb}
\end{figure}

Across all targets and seasons, the $\text{BMC}_2$ has the best mean out-of-sample log score of $-3.02$, though it only marginally outperformed the BLP and LP, which have mean out-of-sample log scores of $-3.03$ and $-3.06$, respectively (\autoref{fig:ls_comb} panel (b)). Across all three test seasons, the BLP has the best mean out-of-sample log scores for the 1 week ahead horizon (\autoref{fig:ls_comb} panel (a)). The $\text{BMC}_2$, which is the most flexible method in this study, has the best mean out-of-sample log scores of $-3.19$ and $-3.34$ for 3 and 4 week ahead horizons, respectively. It also performed as well as the BLP, which also has the best mean out-of-sample log score of $-2.95$ for the 2 week ahead horizon. Across all four target horizons, the BLP is the most accurate method for the 2016/2017 season, while the $\text{BMC}_2$ is the most accurate for the 2017/2018 and 2018/2019 season. 

\begin{figure}[ht]
\includegraphics[scale=0.85]{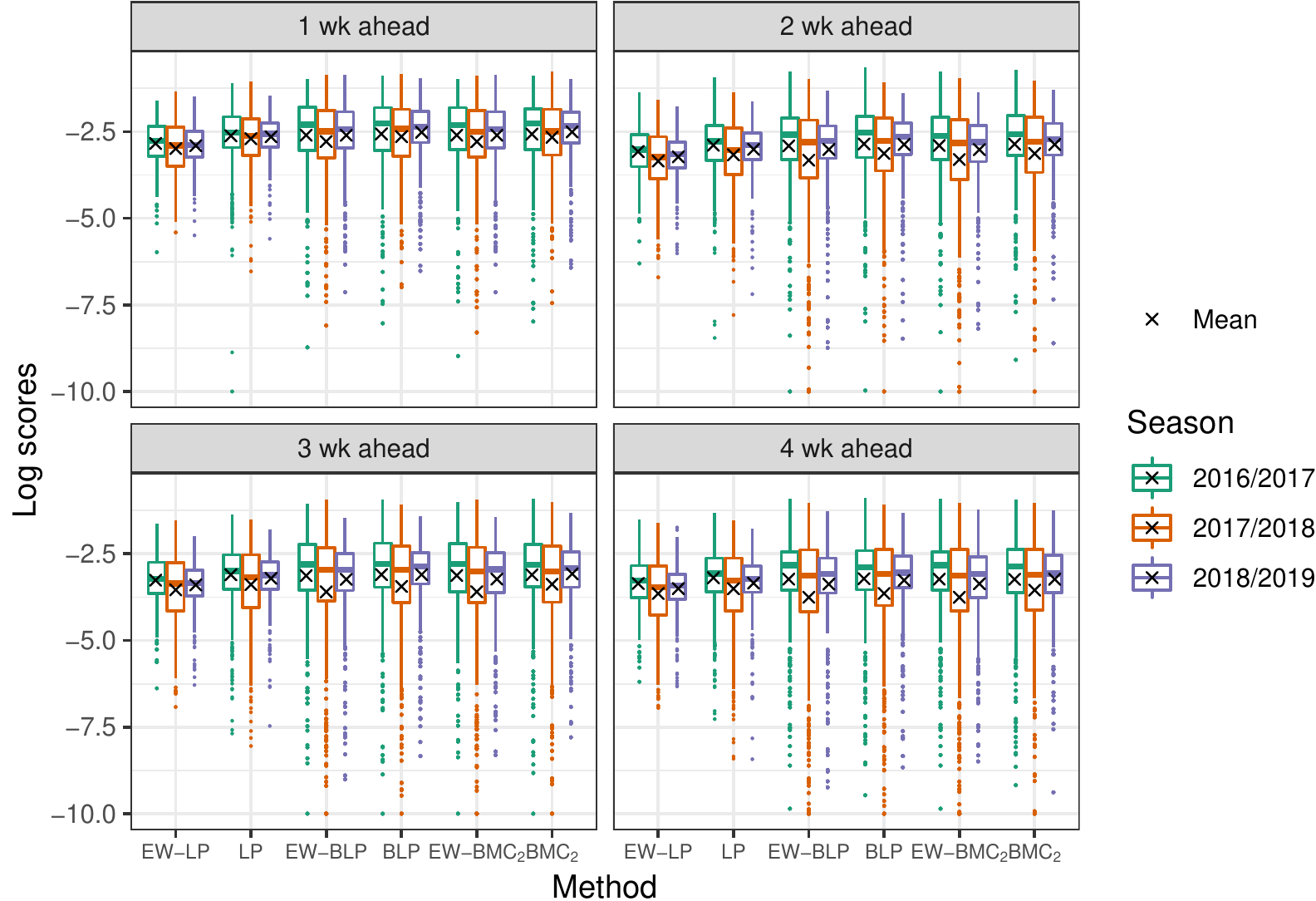}
\caption{Boxplots of out-of-sample, observation-level log scores of ensemble forecasts of wLIL in the U.S. by target and season. Each black cross marker represents a mean out-of-sample log score for a particular target and season. }
\label{fig:box}
\end{figure}

The individual observation-level log scores of the ensemble forecasts from the beta-transformed combination methods exhibit higher variation compared to those from the EW-LP and the LP for all targets and test seasons (\autoref{fig:box}). This is in alignment with the theoretical result that the LP tends to produce overdispersed or wider forecasts, resulting in less extreme log scores. The performance of $\text{BMC}_2$ in the test seasons was less consistent compared to its superior performance across all targets and seasons in the training periods, which indicates of some degree of over-fitting; however, it was always among the top two methods in terms of out-of-sample log score across all targets and seasons (\autoref{fig:ls_comb}). We also see notably higher variation in log scores across all methods for all targets in the 2017/2018 season, which was one of the most severe and longest flu seasons in the recent years \citep{cdc1718}. Section 2 in the Supplementary Material \citep{Wattana22} provides detailed results of mean out-of-sample log scores aggregated and disaggregated at different levels.

\subsubsection{Comparison of combination methods' calibration}

The empirical CDF curves of PIT values from probabilistically calibrated forecasts should follow the CDF of a standard uniform distribution, that is, a diagonal line between 0 and 1. To quantify the deviation from the CDF of a standard uniform distribution, we computed the Cramer distances between the CDF curves of PIT values and the CDF of a standard uniform distribution. Overall, the BLP and $\text{BMC}_2$ methods are more probabilistically calibrated than other methods for the 1 week ahead horizon, as their empirical CDF curves are less deviated from the reference line (\autoref{fig:reli}) and their Cramer distances are the lowest (\autoref{fig:cd1}). However, the forecasts produced from all beta-transformed methods became less calibrated as forecast horizons increased. Across all horizons in the test period, except the 1 week ahead horizon, the empirical CDF curves of the PIT values from the forecasts from the $\text{EW-BMC}_2$ were the most miscalibrated among all beta-transformed methods as indicated by its Cramer distances being the highest. 

Based on the probability plots in \autoref{fig:reli}, all combination methods produced forecasts that lack probabilistic calibration in the test period. Recall that according to theory, the LP will produce ensemble forecasts with too wide predictive distributions when individual models are well-calibrated \citep{gneiting2013}. The results (\autoref{fig:reli}) in our application during the training period are consistent with this theory. Specifically, the forecasts from the LP and EW-LP tended to be to wide, i.e., more observed values concentrated near the center of predictive distributions than expected for a well-calibrated model as indicated by the slopes of PIT CDF curves being higher than 1 for intermediate PIT values and lower than 1 near 0 and 1 (\autoref{fig:reli}). 

The beta transformed combination methods were successful at correcting the overdispersion of the LP and EW-LP. However, their forecasts are miscalibrated in the test period due to under-prediction (\autoref{fig:reli} panel (b)). The empirical CDF curves of the PIT values from the BLP, EW-BLP, $\text{BMC}_2$, and $\text{EW-BMC}_2$ were below the reference line across all PIT values from the test period, indicating consistent under-prediction, i.e., the predictive distributions tended to be concentrated below the observed values. 

Despite the under-prediction across horizons observed in both training and test periods, the beta-transformed combination methods probabilistic calibration was notably better in the training period, especially for 3 and 4 week ahead horizons (\autoref{fig:reli} panel (a)). In addition, the calibration of the forecasts produced from the EW-BLP and $\text{EW-BMC}_2$ was similar to that of the forecasts produced from the BLP and $\text{BMC}_2$ in the training period, but more miscalibrated in the test period, as their Cramer distances are notably higher than those of the BLP and $\text{BMC}_2$ in the test period. Similar to the calibration observed in the test period, The LP and EW-LP also produced wide forecasts in the training period, albeit with more miscalibration in the upper tails across all horizons. 

The probability plots by target-season pairs (Figure S6, Supplemental Material \citep{Wattana22}) show similar calibration results as in \autoref{fig:reli} \textemdash the empirical CDF curves of the PIT values of forecasts produced from the LP methods in the test seasons also appear miscalibrated in the lower tail. The calibration of all methods by target-season pairs are discussed in more details in Section 3 in Supplemental Material \citep{Wattana22}. 

\subsubsection{Comparison of accuracy and calibration between combination methods with equally weighted and with optimally weighted individual models}

\begin{figure}[htp]
\includegraphics[scale=0.68]{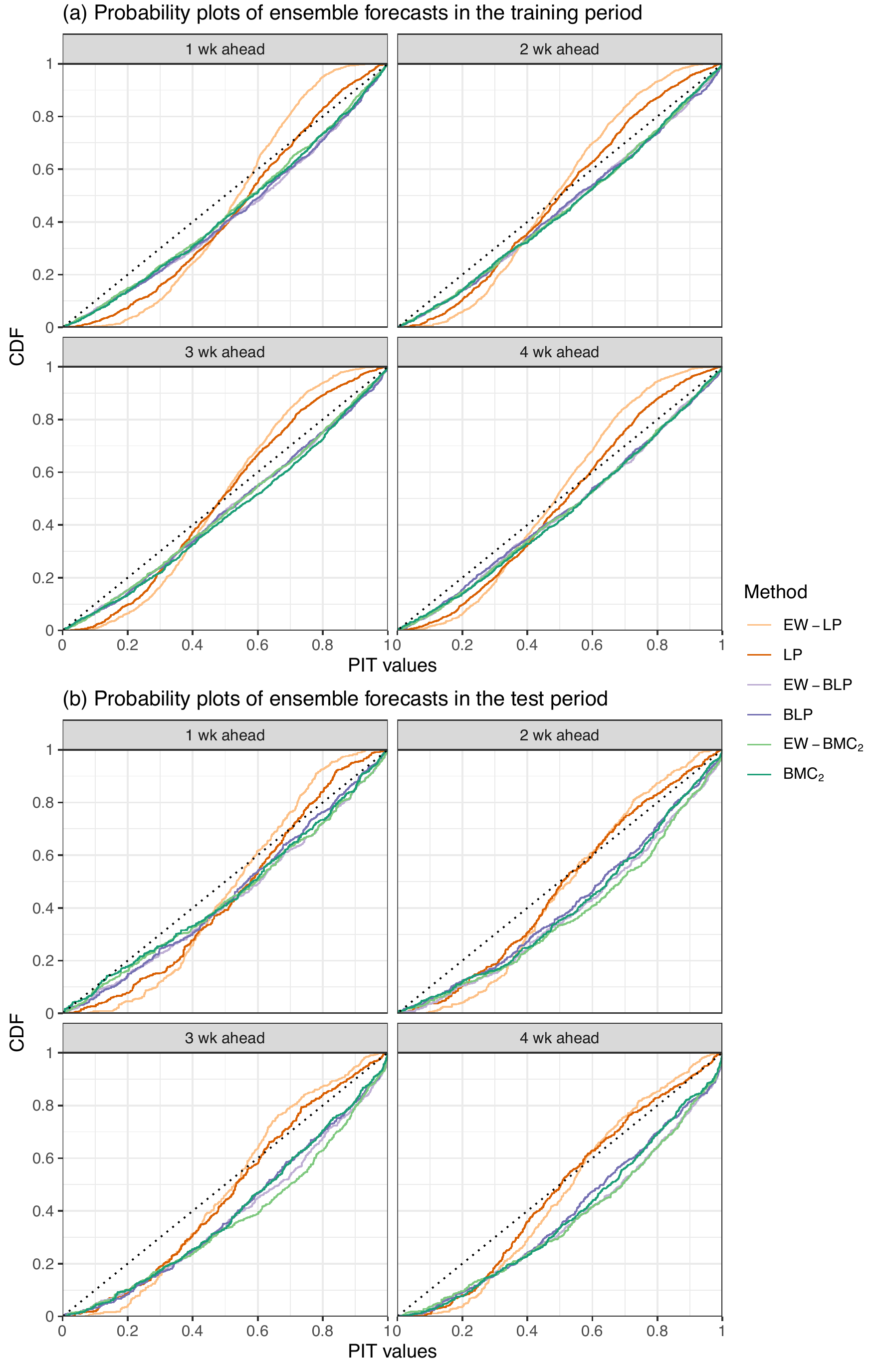}
\caption{Probability plots show the empirical CDF curves of PIT values of ensemble forecasts by target. Panel (a) shows probability plots of ensemble forecasts in the training period. Panel (b) shows probability plots of ensemble forecasts by target in the test period (the 2016/2017 to 2018/2019 season). The black diagonal dashed line is the reference line for assessing probabilistic calibration. The more an empirical CDF curve of the PIT values deviates from the reference line, the more miscalibrated the forecasts produced from the corresponding method is. The shape of the curves can be used to diagnose the cause of miscalibration.}
\label{fig:reli}
\end{figure}

\begin{figure}[ht]
\includegraphics[scale=0.62]{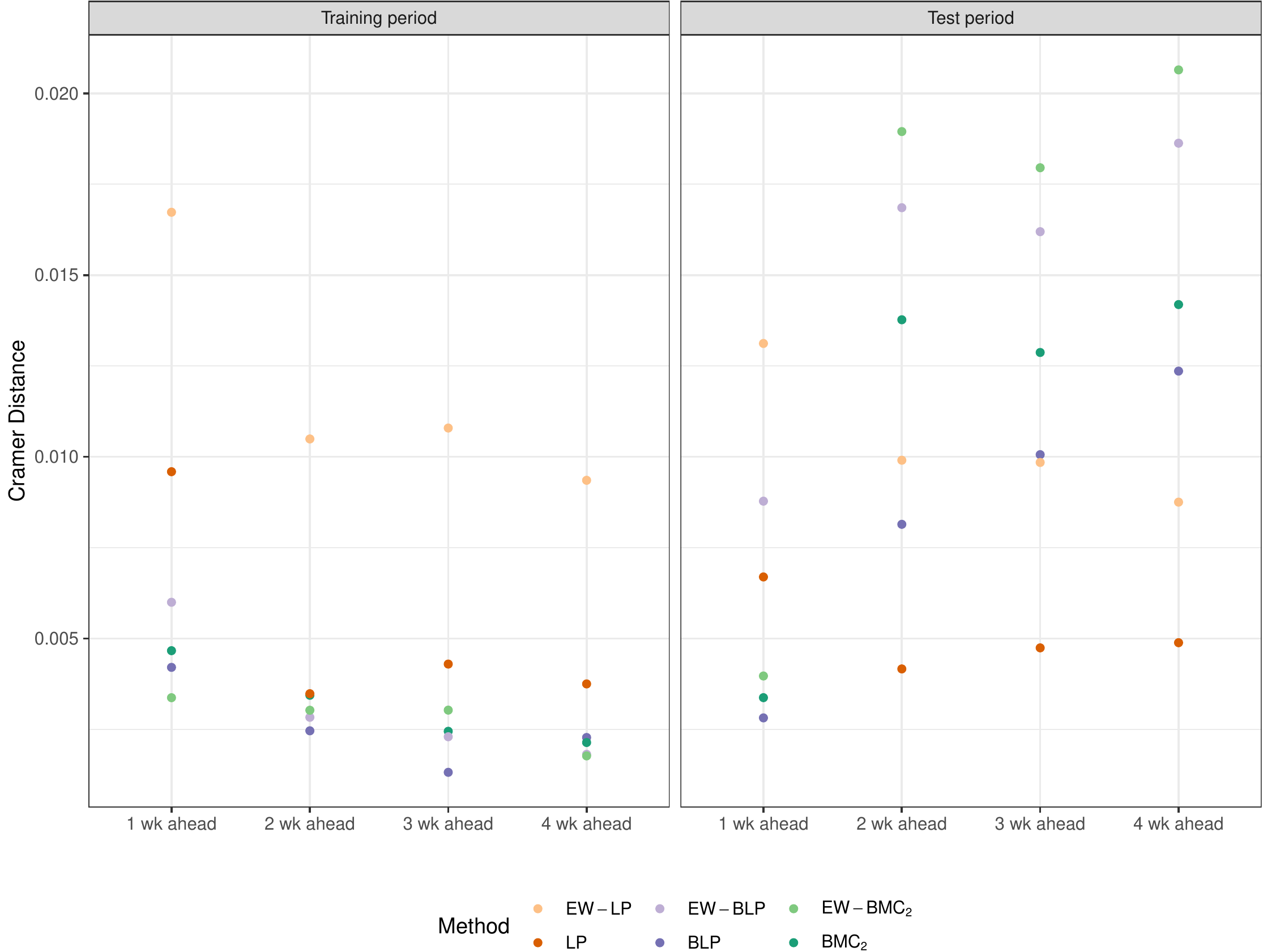}
\caption{Cramer distances between empirical CDF curves of PIT values and the reference line in \autoref{fig:reli}. Cramer distances can be considered a quantitative summary of the probabilistic calibration. The higher the Cramer distance between an empirical CDF curve and the reference line is, the more miscalibrated the forecasts produced from the corresponding combination method is.}
\label{fig:cd1}
\end{figure}

The equally weighted variations of the combination methods (EW-LP, EW-BLP, and $\text{EW-BMC}_2$), though more parsimonious, had sub-optimal forecast accuracy compared to their counterparts that assigned weights to the individual models in this application. The EW-LP was the worst overall method across all targets and seasons, and all equally weighted variations had worse mean out-of-sample log scores compared to their more complex counterparts (\autoref{fig:ls_comb}). Additionally, the equally weighted ensembles generally had poorer calibration than the corresponding weighted variations in both the training period and the test period (Figures \ref{fig:reli}, \ref{fig:cd1}).

\section{Discussion}\label{4}

As demonstrated in the forecasting literature, in many settings ensemble forecasts have consistent superior performance and give decision makers the ability to unify the strengths and diversity of individual models into one forecast. These particular advantages are of great importance in practice for infectious disease forecasting \citep{defelice2017,reich2019,chowell2020,mcandrew2020,ray2020,ray_2022}. This work aims to offer insight into forecast accuracy and calibration of parametric combination methods in which calibration and individual model weight estimation happen simultaneously in the application of seasonal influenza forecasting in the U.S. 

We applied the linear pool, beta-transformed linear pool, and the finite beta mixture combination method to available forecasts in the FluSight challenge to produce ensemble forecasts for seasonal influenza in the U.S. retrospectively for three test seasons and compared their performance. Our results showed that two of the combination methods included in this study, the BLP and $\text{BMC}_2$, offered consistently superior forecast accuracy relative to the LP and EW-LP. Either the BLP or the $\text{BMC}_2$ or both delivered better mean log scores across the test seasons compared to the other methods for all targets and across all targets for all seasons. Despite using different methods to create ensemble forecasts, our findings are in agreement with the findings in \cite{rumack_21} that combination methods that take into account forecast calibration improve accuracy of seasonal influenza ensemble forecasts in the U.S. 

The $\text{BMC}_2$ uses twice as many parameters as the BLP, but only marginally outperformed it in two out of four targets and five out of twelve target-season pairs. Considering the large number of individual models in the FluSight challenge, the BLP may be more easily applicable in practice compared to the $\text{BMC}_2$ as it has half as many individual model weights and beta parameters to estimate. Although the LP under-performed relative to the BLP and $\text{BMC}_2$ for most targets and seasons, the differences in mean log scores were typically small. More parsimonious combination methods with fixed, equally weighted individual model weights, namely the EW-LP, EW-BLP and $\text{EW-BMC}_2$, appear to not be flexible enough to deliver superior performance compared to the other methods in this study. While this is the case for this application, combination methods using equal weights with or without the beta transformation could be useful in other applications where it might be difficult to estimate individual model weights when available models change over time or training data are limited \citep{ray_2022}.

The results on the probabilistic calibration of the ensemble forecasts measured by the uniformity of the PIT values are less straightforward. The results for the LP and EW-LP forecasts indicate that they had too wide predictive distributions across all targets. Despite the beta-transformed combination methods' success at correcting the overdispersion of the ensemble forecasts from the LP and EW-LP, their forecasts exhibited a pattern of systematic under-prediction. This under-prediction is relatively more pronounced at longer forecast horizons, especially in the 2017/2018 and 2018/2019 influenza seasons. In the 2017/2018 season, which was a large influenza season in the U.S. \citep{cdc1718}, the ensemble forecasts from all combination methods under-predict to some extent. Note that an overdispersed forecaster, such as the LP and EW-LP, has the advantage of being more likely to capture an extreme season, though it may not be optimal overall based  on proper scores. These more conservative methods may be desirable in applications in which stakeholders want to avoid missing a large season at the expense of having too wide forecasts.

Due to the flexibility of the beta-transformed combination methods and season-to-season variations of influenza, the lack of calibration could stem from overfitting to the training season data as the CDFs of PIT values in training seasons showed better calibration overall.  In this study, the parameters of the beta-transformed combination methods were estimated by maximum likelihood, which is not equivalent to a measure of probabilistic calibration. If probabilistic calibration is of critical importance for the application at hand, other approaches such as post-hoc calibration techniques that directly target a calibrated forecast distribution may be appropriate. 

Exploring different approaches to select training periods can be useful. For instance, \cite{baran2018} use rolling training periods in the application of a similar set of combination methods to wind speed and precipitation forecasting and \cite{rumack_21} constructs a training period that takes into account seasonality of epidemic forecasting. In addition, combination methods that require the joint estimation of all parameters, including the parameters in the individual models, may be considered in a setting where the underlying model structure of individual models are known. An example of one of these methods is the mixture EMOS model proposed by \cite{baran2016}. Since the FluSight challenge provides forecasts only, in this work we selected combination methods that do not require reproduction of forecasts from individual models.

For combining forecasts in outbreak settings, the beta-transformed linear pool (BLP) is a promising alternative to standard linear pooling (LP) methods. Compared with LP,  the BLP has only two additional parameters, $\alpha$ and $\beta$, and the simple modification of the log likelihood function makes the BLP applicable to combining forecasts in a binned probability format. The $\text{BMC}_2$ may add value in instances where the BLP is not flexible enough, though we only see marginal improvement in the mean out-of-sample log scores in our application. 

As infectious disease forecasting has come to the forefront of the public health effort in formulating well-informed policies in response to outbreaks, it is critical to gain insight on model combination approaches in order to combine individual models' strengths and to produce accurate ensemble forecasts. This study demonstrates an effort to improve our understanding of how forecast combination methods compare in a setting of an infectious disease with well-established surveillance data pipeline like seasonal influenza. 
\begin{funding}
This work has been supported by the National Institutes of General Medical Sciences (R35GM119582) and the US CDC (1U01IP001122). The content is solely the responsibility of the authors and does not necessarily represent the official views of NIGMS, the National Institutes of Health, or CDC.
\end{funding}
\begin{supplement}
\stitle{Individual Forecasting Model Information and Additional Results}
\sdescription{This supplement contains sections 1–3 including individual forecasting model information and additional results of the application.}
\end{supplement}

\bibliographystyle{imsart-nameyear} 
\bibliography{main.bib}

\end{document}

\title{Invidual Forecasting Model Information and Additional Results}

\section{Individual forecasting models}\label{s1}

Prospective forecasts of 1-4 week ahead wILI for the United States and the 10 Health and Human Services (HHS) regions from 27 models in \autoref{tab:modlist} are available in the FluSight Network repository \citep{flusight}. 

\begin{table}[H]
\caption{List of individual forecasting models in the FluSight Network repository}\label{tab:modlist}
\centering
\setlength{\tabcolsep}{4pt} 
\begin{tabular}{p{1.6cm} l p{9cm} }
\hline
Team     & Model Abbreviation & Model Description  \\ 
\hline
CU       & EAKFC\_SEIRS       & Ensemble Adjustment Kalman Filter SEIRS \citep{Pei2017} \\ 
~        & EAKFC\_SIRS       & Ensemble Adjustment Kalman Filter SIRS \citep{Pei2017}\\
~        & EKF\_SEIRS        & Ensemble Kalman Filter SEIRS \citep{Yang2014}\\
~        & EKF\_SIRS         & Ensemble Kalman Filter SIRS \citep{Yang2014}\\
~        & RHF\_SEIRS        & Rank Histogram Filter SEIRS \citep{Yang2014}\\
~        & RHF\_SIRS          & Rank Histogram Filter SIRS \citep{Yang2014}\\
~        & BMA                & Bayesian Model Averaging \citep{yamana2017}\\
\hline
Delphi   & BasisRegression    & Basis Regression ({\tt epiforecast} defaults) \citep{brooks2015}\\ 
~        & DeltaDensity1      & Delta Density ({\tt epiforecast} defaults) \citep{brooks2018}\\
~        & DeltaDensity2     & Markovian Delta Density ({\tt epiforecast} defaults) \citep{brooks2018}\\ 
~        & EmpiricalFuture    & Empirical Futures ({\tt epiforecast} defaults) \citep{brooks2015}\\ 
~        & EmpiricalTraj      & Empirical Trajectories ({\tt epiforecast} defaults) \citep{brooks2015}\\ 
~        & Uniform            & Uniform Distribution \cite{flusight}\\ 
\hline
LANL     & DBMplus                & Dynamic Bayesian SIR Model with discrepancy \citep{Osthus2019}\\ 
\hline
ReichLab & KCDE               & Kernel Conditional Density Estimation \citep{ray2017} \\ 
~        & KCDE backfill      & Kernel Conditional Density Estimation with backfill \citep{ray2017}\\ 
~        & KDE                & Kernel Density Estimation and penalized splines \citep{ray2018_ens}\\ 
~        & SARIMA1            & SARIMA model without seasonal differencing \citep{ray2018_ens}\\ 
~        & SARIMA2            & SARIMA model with seasonal differencing \citep{ray2018_ens}\\ 
\hline
FluOutlook & Mech      & Mechanistic GLEAM Ensemble\citep{flusight}\\
~          & MechAug & Augmented Mechanistic GLEAM Ensemble \citep{flusight}\\
\hline
Protea & Cheetah         & Ensemble of dynamic harmonic model and historical averages \citep{flusight}\\ 
~      & Kudu            & Subtype weighted historical average model \citep{flusight}\\ 
~      & Springbok       & Dynamic Harmonic Model with ARIMA errors \citep{flusight}\\ 
\hline
FluX & ARLR              & Auto Regressive model with Likelihood Ratio based Model Selection \citep{flusight} \\ 
~    & LSTM              & Recurrent Neural Network (Long Short-Term Memory) \citep{flusight}\\
\hline
UA        & EpiCos           & Epidemic Cosine with Variational Data Assimilation \citep{flusight}\\ 
\end{tabular} 
\end{table}

\section{Mean out-of-sample log scores at different aggregation levels} \label{s2}

\begin{figure}[t]
\includegraphics[scale=0.62]{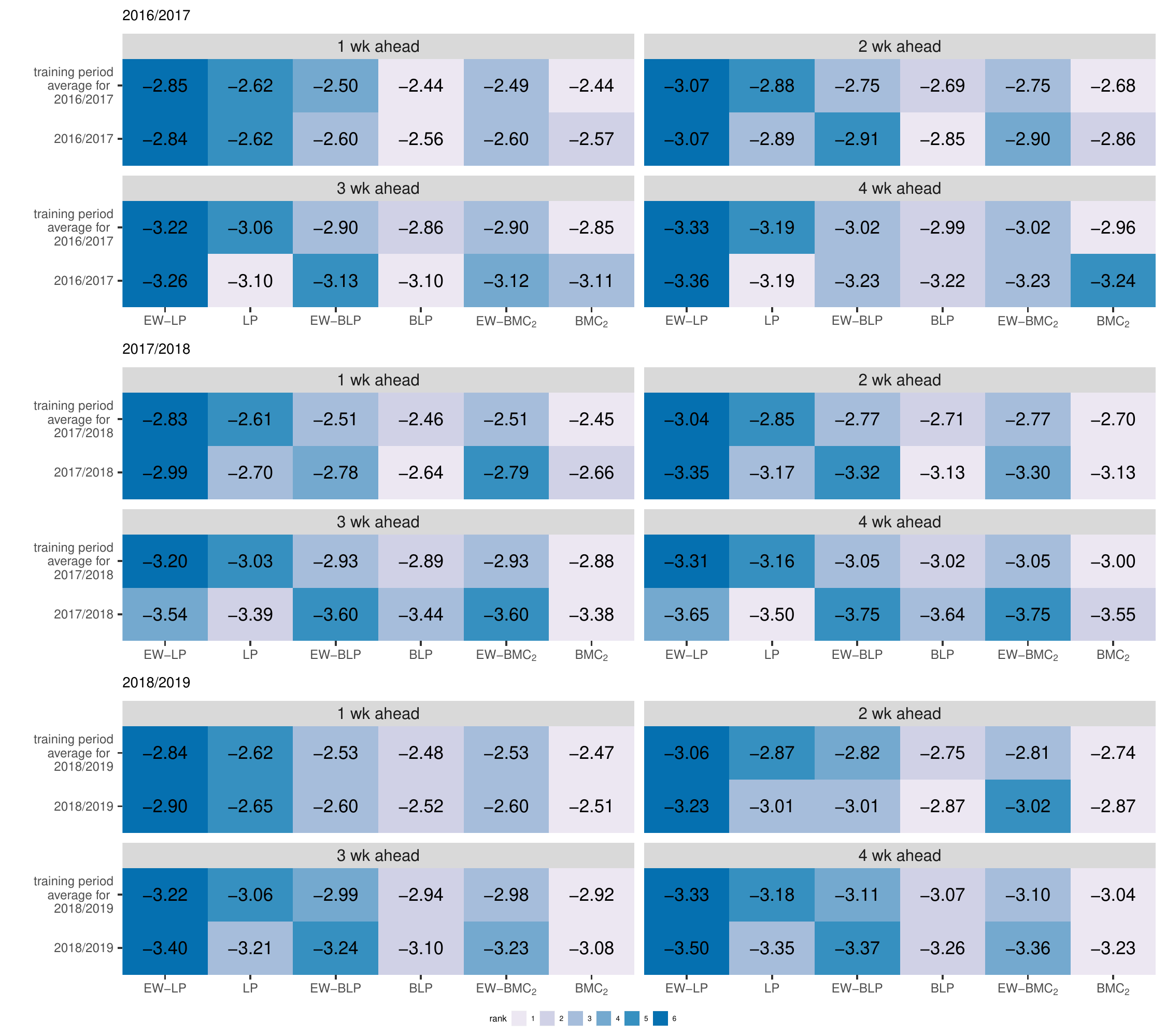}
\caption{Mean training and out-of-sample log scores for the 1-4 week ahead targets in the 2016/2017, 2017/2018, and 2018/2019 season}
\label{fig:lssup}
\end{figure}

Forecast accuracy of ensemble forecasts produced from all combination methods measured by unaggregated mean out-of-sample log scores varied across targets in the three test seasons (\autoref{fig:lssup}). For 8 out of 12 target-season pairs, the $\text{BMC}_2$ and BLP were the two top performing methods. Specifically, both the BLP and $\text{BMC}_2$ had the best mean out-of-sample log scores for the 2 week ahead horizon in the 2017/2018 and 2018/2019 seasons. The BLP outperformed other methods for 1 and 3 week ahead horizons in the 2016/2017 season and for 1-2 week ahead horizons in the 2016/2017 season. The $\text{BMC}_2$ outperformed the other four methods for the 3 week ahead horizon in the 2017/2018 season and for 1,3, and 4 week ahead horizons in the 2018/2019 season. The LP had the best score for the 3-4 week ahead horizons in the 2016/2017 season and the 4 week ahead horizon in the 2017/2018 season. The EW-LP was the worst performing method for most target and season pairs, while the EW-BLP, and $\text{EW-BMC}_2$ yielded the worst log scores for the 3 and 4 week ahead targets in the 2017/2018 season.

\begin{figure}[H]
\centering
\includegraphics[scale=0.8]{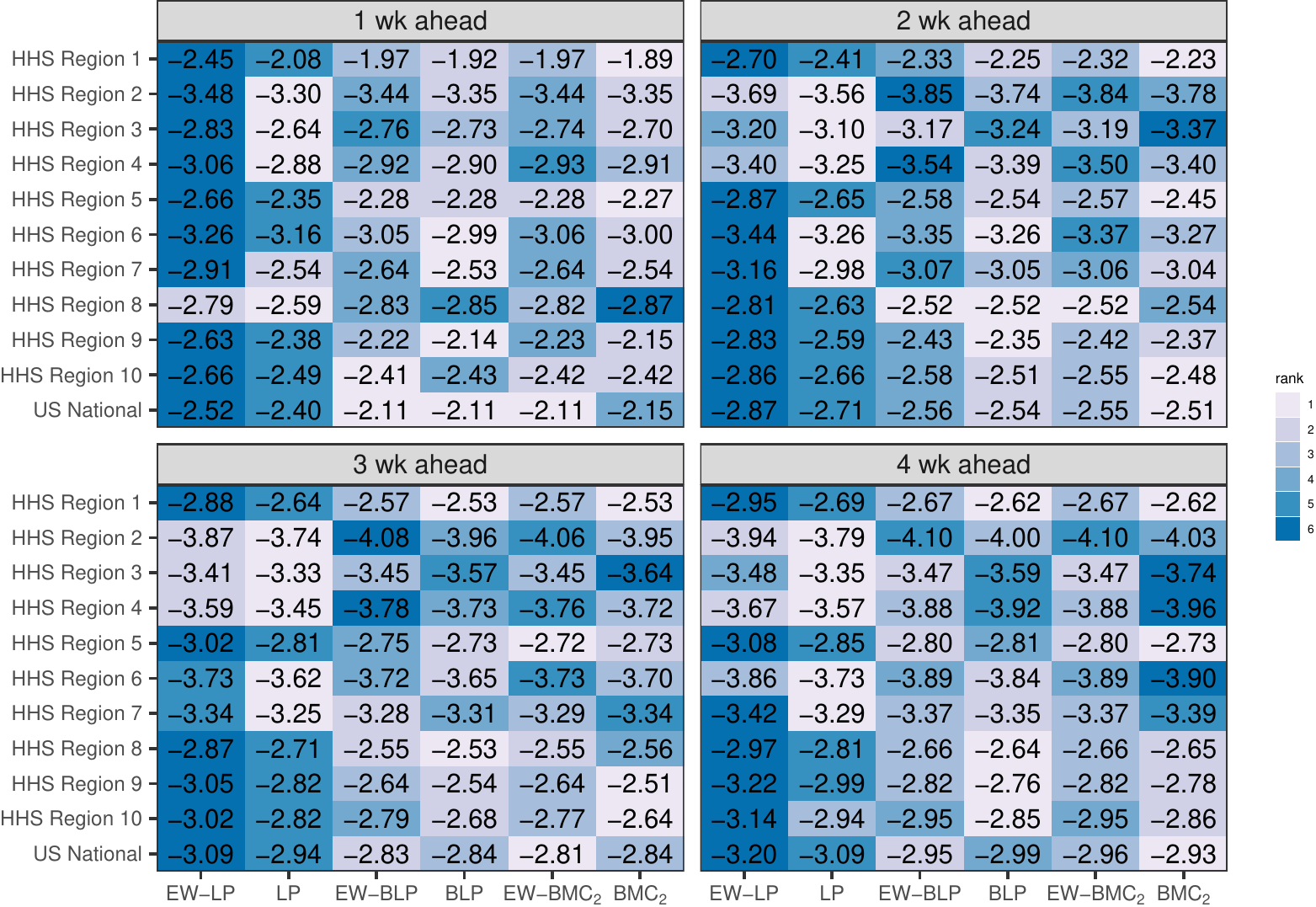}
\caption{Mean log score across all weeks by HHS region in the 2016/2017 influenza season}
\label{fig:lsloc1}
\end{figure}

\begin{figure}[H]
\centering
\includegraphics[scale=0.8]{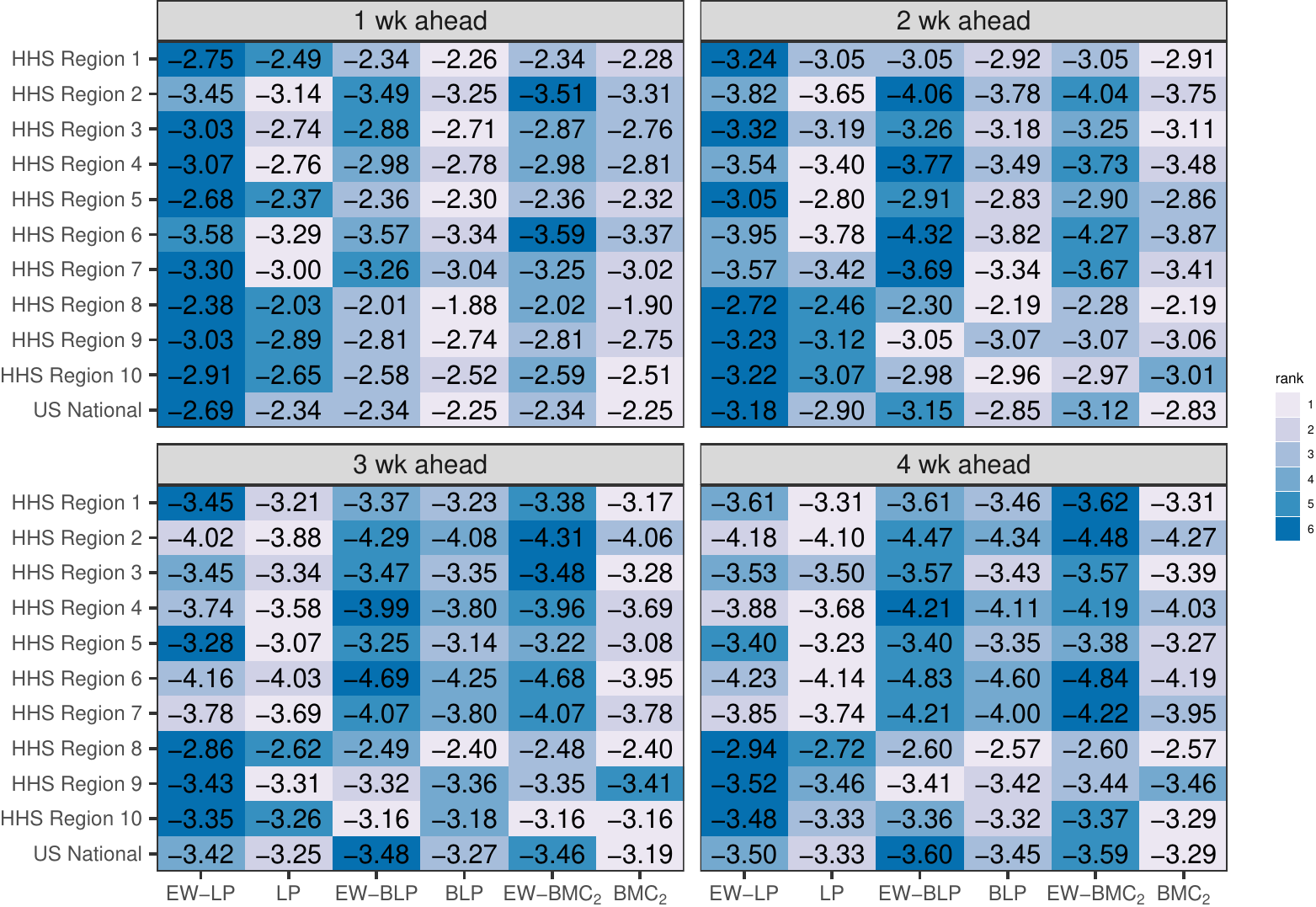}
\caption{Mean log score across all weeks by HHS region in the 2017/2018 influenza season}
\label{fig:lsloc2}
\end{figure}

\begin{figure}[H]
\centering
\includegraphics[scale=0.8]{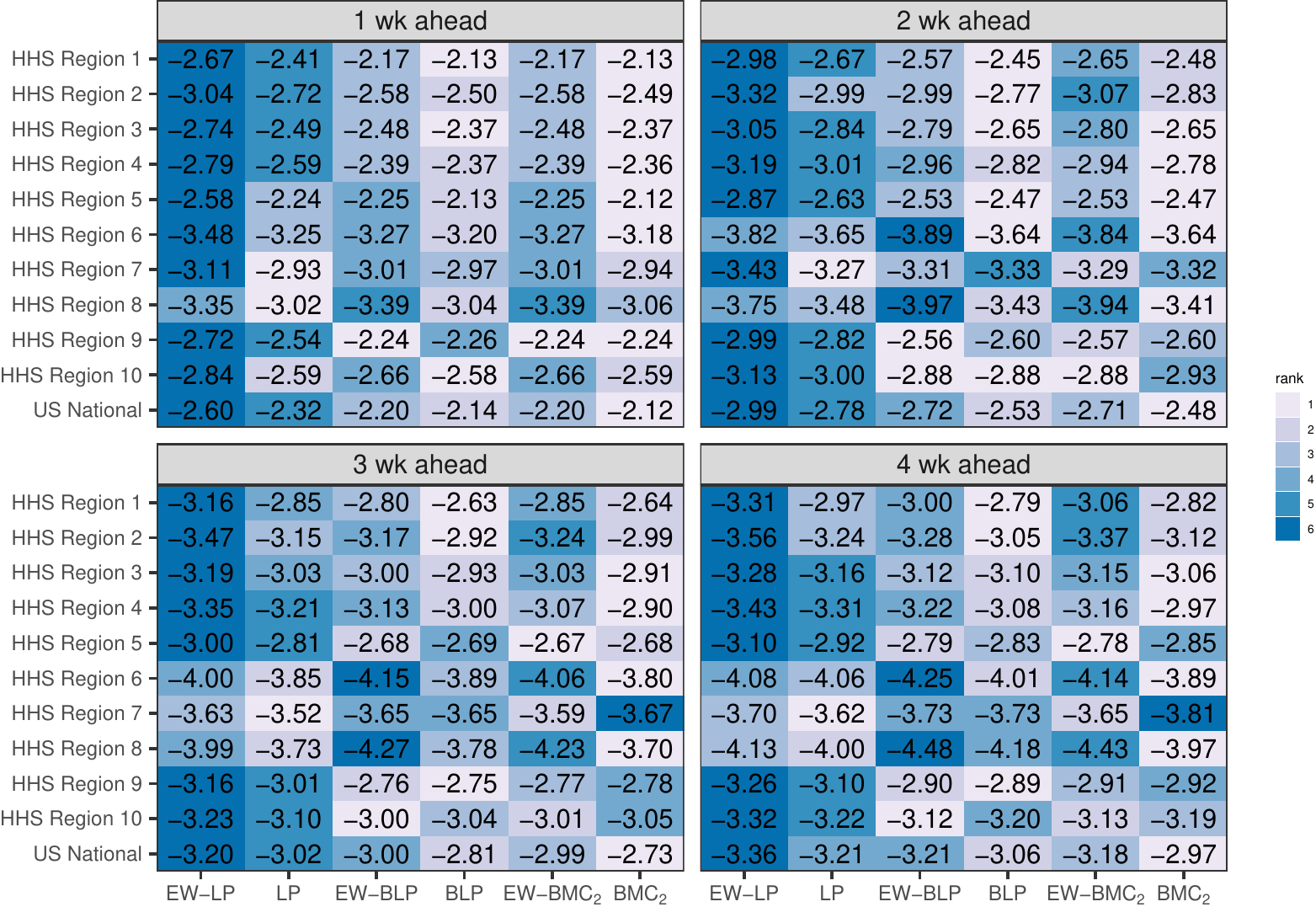}
\caption{Mean log score across all weeks by HHS region in the 2018/2019 influenza season}
\label{fig:lsloc3}
\end{figure}

For most locations, the EW-LP was also the worst performing method across all targets and seasons (see \autoref{fig:lsloc1}, \autoref{fig:lsloc2}, and \autoref{fig:lsloc3}). Either the BLP or $\text{BMC}_2$ or both were the two top performing methods across all targets for about half of the locations in the 2016/2017 and 2017/2018 seasons and for most locations in 2018/2019 season. The LP was one of the top two performing method for the 3 and 4 week ahead targets for most locations in the 2017/2018 season. These variations in out-of-sample log scores were also shown in Figure 4 in the main article.


\section{Probability plots of ensemble forecasts by target-season pair} \label{A3}

\begin{figure}[htp]
\centering
\includegraphics[scale=0.6]{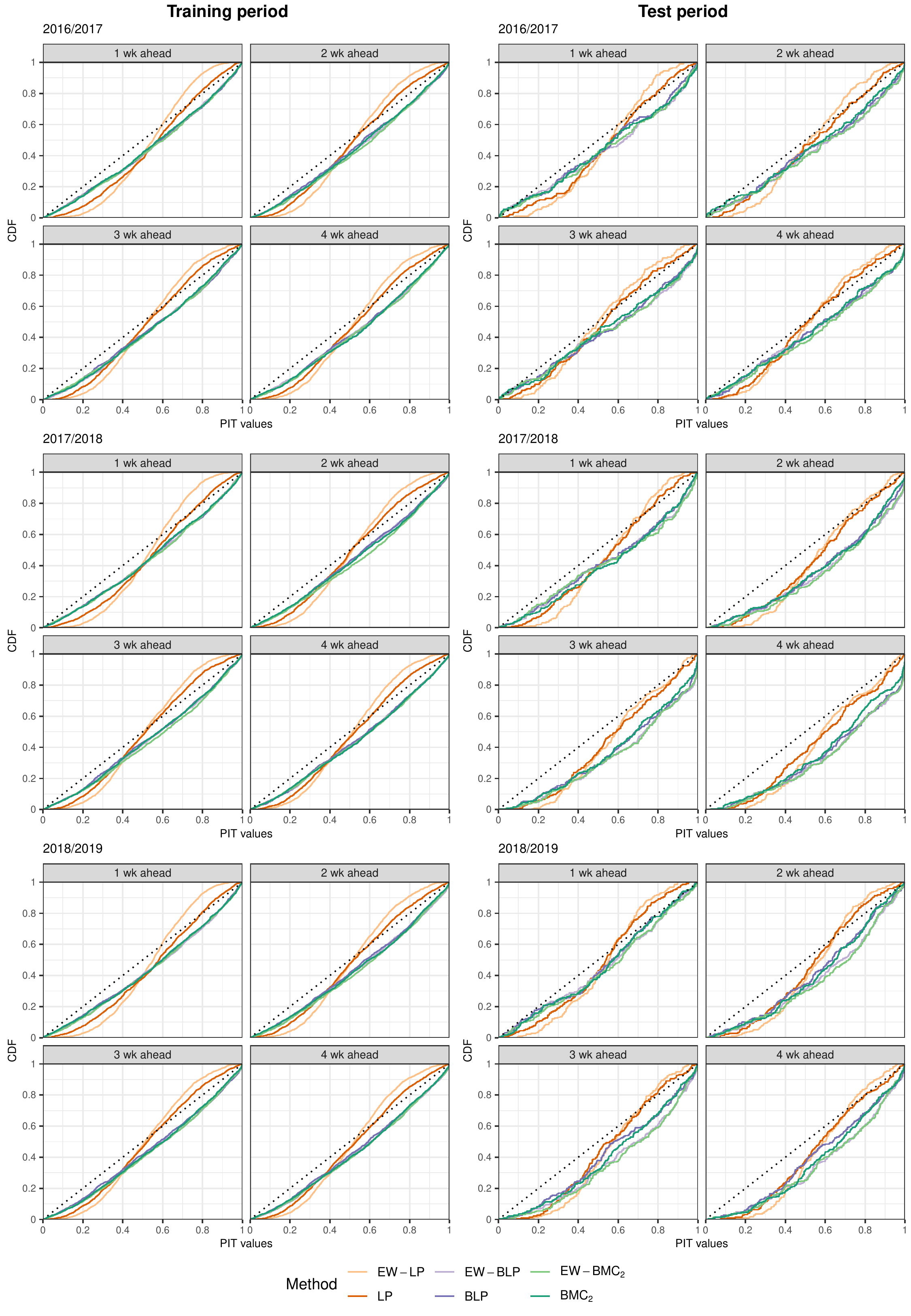}
\caption{Probability plots by target and season. The black diagonal line is the reference line for assessing probabilistic calibration. The more an empirical CDF curve deviates from the reference line, the more miscalibrated the forecasts produced from the corresponding combination method is.}
\label{fig:reli_all}
\end{figure}

\begin{figure}[htp]
\centering
\includegraphics[scale=0.58]{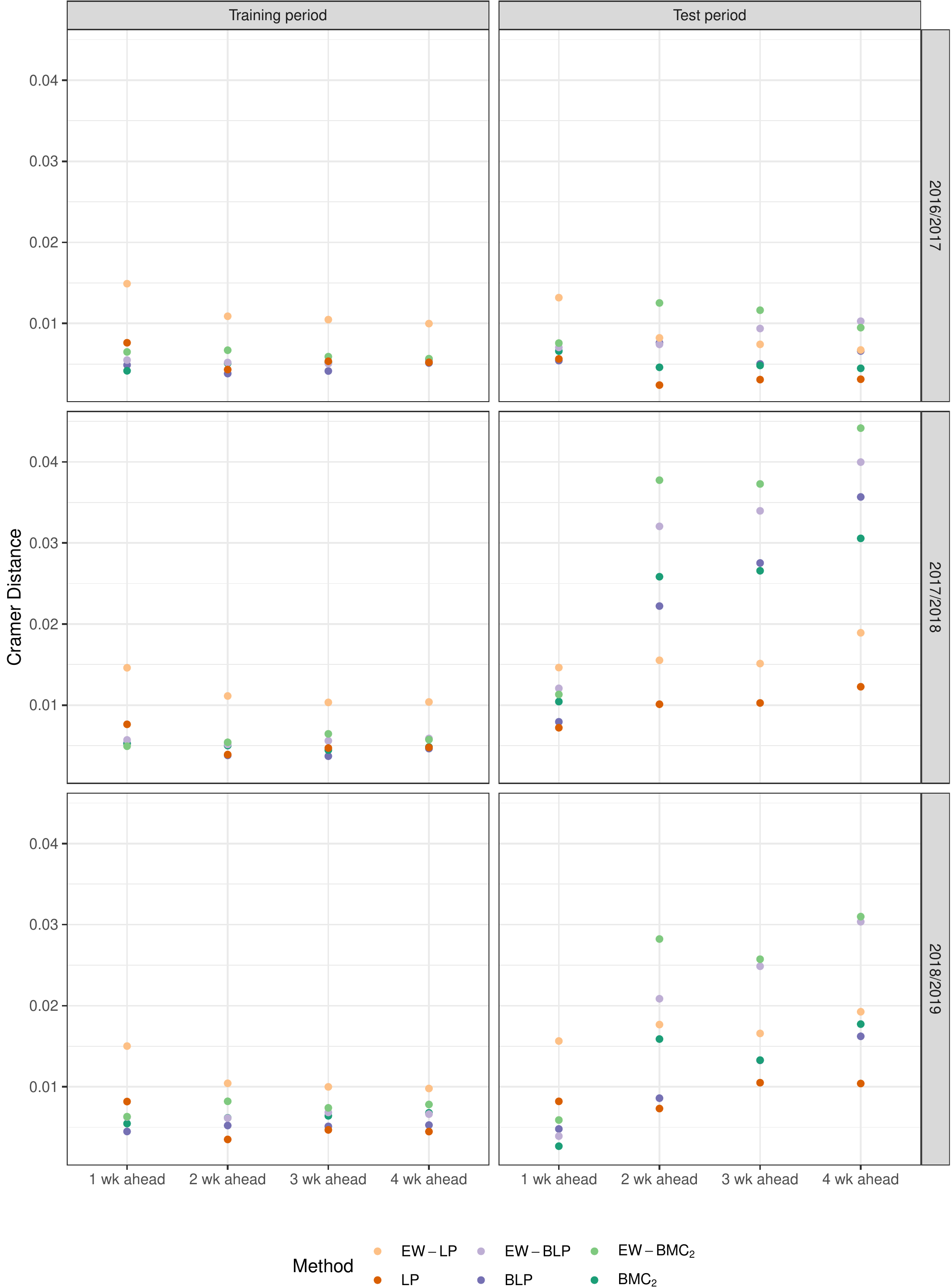}
\caption{Cramer distances between empirical CDF curves of PIT values and the reference line in \autoref{fig:reli_all}. The higher the Cramer distance between an empirical CDF curve and the reference line is, the more miscalibrated the forecasts produced from the corresponding combination method is.}
\label{fig:cd2}
\end{figure}

 \autoref{fig:reli_all} and \autoref{fig:cd2} highlight season-to-season variations in the probabilistic calibration of the ensemble forecasts. For all targets, the out-of-sample forecasts from the EW-LP and LP methods were slightly too wide in the 2016/2017 season as the distributions of PIT values were concentrated around intermediate PIT values. However, the probability plots indicated under-prediction for 2 to 4 week ahead horizons in the 2017/2018 seasons as too few forecasts had PIT values below approximately 0.6. In the 2018/2019 season, they produced too wide forecasts for 1 to 2 week ahead horizons, while under-prediction drove their miscalibration for 3 to 4 week ahead horizons. 
 
 Out-of-sample forecasts produced from the beta-transformed combination methods exhibited modest under-prediction for all targets in the 2016/2017 seasons as the empirical CDF curves are below the reference line across all PIT values. The Cramer distances between the empirical CDF curves and the reference line indicated they deviated farther from the reference line compared to the LP's forecasts. Their 2 to 4 week ahead forecasts were more miscalibrated in the 2017/2018 season, as shown in both \autoref{fig:reli_all} and \autoref{fig:cd2}. 
 
 The beta-transformed combination methods yielded better calibrated out-of-sample forecasts compared to the EW-LP and LP for the 1 week ahead target, but showed substantial under-prediction for the rest of the targets in the 2018/2019 season. This is in alignment with the results of summary measure of probabilistic calibration by Cramer distances, as the Cramer distances between the empirical CDF curves of the PIT values of beta-transformed combination methods and the reference line are lower than Cramer distances between the empirical CDF curves of the PIT values of the EW-LP and LP and the reference line for the 1 week ahead horizon, but higher for all other horizons.
 
 The lack of calibration of the ensemble forecasts generated from the beta-transformed combination methods for most targets in the test seasons, evident in \autoref{fig:reli_all} and \autoref{fig:cd2}, was in contrast with modest under-prediction and lower Cramer distances from the reference line shown in the training period. In the 2017/2018 season, which was a large influenza season in the U.S., the ensemble forecasts from all combination methods appeared to under-predict. 

\bibliographystyle{imsart-nameyear} 
\bibliography{supplement.bib}